\documentclass[prl,aps,twocolumn,floatfix,showpacs]{revtex4}%
\usepackage{amsfonts}
\usepackage{amsmath}
\usepackage{amssymb}
\usepackage{graphicx}%
\providecommand{\U}[1]{\protect\rule{.1in}{.1in}}
\begin{document}
\title{Radio-frequency magnetic response of vortex lattices undergoing structural transformations}
\author{R. Prozorov}
\email{prozorov@ameslab.gov}
\affiliation{Ames Laboratory and Department of Physics and Astronomy, Iowa State
University, Ames IA 50011, USA}
\author{V.~G. Kogan}
\affiliation{Ames Laboratory and Department of Physics and Astronomy, Iowa State
University, Ames IA 50011, USA}
\author{M. D. Vannette}
\affiliation{Ames Laboratory and Department of Physics and Astronomy, Iowa State
University, Ames IA 50011, USA}
\author{S.~L. Bud'ko}
\affiliation{Ames Laboratory and Department of Physics and Astronomy, Iowa State
University, Ames IA 50011, USA}
\author{P.~C. Canfield}
\affiliation{Ames Laboratory and Department of Physics and Astronomy, Iowa State
University, Ames IA 50011, USA}
\date{21 February 2007}


\begin{abstract}
We report a clear anomaly in the rf response of a tunnel-diode resonator
measured in borocarbide crystals in dc fields along the $c-$axis. We associate
the anomaly in the dynamic magnetic susceptibility, $\chi$, with a structural
transition in the vortex lattice. $\chi$ is sensitive to the Campbell
penetration depth $\lambda_{C}$ of the rf perturbation into the mixed state.
$\lambda_{C}$ depends on the vortex lattice elastic moduli, which in turn
depend on the vortex lattice structure. The high-field transition line is
clearly mapped and shown to behave in agreement with theory.

\end{abstract}
\maketitle

In the classical picture of type-II superconductors, vortices should form a
hexagonal lattice in isotropic or cubic materials; the same should happen in
tetragonal materials for applied magnetic fields along the crystallographic
$c$ axis. It has been known for a long time that experimentally this is rarely
the case \cite{hueb,Christen}. The reason for such behavior was, to a large
extent, clarified after the nonlocal effects were incorporated into the London
description of intervortex interaction \cite{cubic,boro,Affleck}. This effort
resulted in the realization that the vortex lattice (VL) structure depends on
the Fermi surface, the order parameter symmetry, as well as on the applied
field, temperature, scattering strength, and even on thermal fluctuations of
vortex positions \cite{GK}. Currently, vortex lattices structural phase
transitions are best studied in the rare earth-nickel-borocarbides
\cite{Paul,Nature,ORNL,Grenoble,Argonne,Gammel,Vinn,Sakata,nose}, but in fact
present in many superconductors including cubic Nb or V$_{3}$Si
\cite{Forgan,V3Si,Mona}.

Hence, in addition to boundaries of different dynamic behavior (e.g., liquid,
glass, or the \textquotedblleft Abrikosov crystal"), the phase diagram of the
vortex state contains lines separating different periodic structures within
the domain of the Abrikosov crystal. Information about these structures is
consequential because the VL elastic properties are structure dependent,
whereas the dynamic behavior is affected by the lattice elastic moduli
\cite{Blatter}.

Vortex lattices have been probed by Bitter decoration \cite{Vinn}, Small Angle
Neutron Scattering (SANS) \cite{Paul,Nature,ORNL,Grenoble,Gammel}, and by
Scanning Tunneling Microscopy \cite{Argonne,Sakata}. (Hereafter we are talking
about tetragonal crystals in fields along the $c$-axis). Notwithstanding
considerable progress made in locating the transition lines in low fields (an
example is given in Fig.\thinspace\ref{fig3}), the complete mapping of
transitions had proven elusive mostly due to poor high-field sensitivity of
these techniques. Calculations based on the Ginzburg-Landau approach
\cite{Argonne} suggested that the transition line for transformation of
rhombic (triangular) to square (R-S) structures may end up at the upper
critical field $H_{c2}(T)$ . However the SANS data on LuNi$_{2}$B$_{2}$C have
shown the transition line bending away of $H_{c2}(T)$ \cite{nose}. Gurevich
and Kogan \cite{GK} suggested that softening of the lattice near $H_{c2}(T)$
causes the vortex fluctuations to increase thus smearing a weak
square-symmetric intervortex interaction and making it isotropic (see also
\cite{Dorsey}). The model predicts that in small fields the R-S transition
line $H_{\Box}(T)$ slowly increases with temperature but then sharply turns
upward to avoid $H_{c2}(T)$ and become a double-valued function.

The dynamic magnetic susceptibility $\chi$ of a superconductor in the mixed
state can be measured by the tunnel-diode resonator technique described
elsewhere \cite{prozorov2000,prozorov2000a}. In short, the frequency shift,
$\Delta f$, due to a small change in the resonator induction caused by the
finite penetration of the ac field into the sample in the mixed state is
recorded. This shift is proportional to the change in the total rf penetration
depth, $\delta\propto\Delta f$ as long as $\delta$ is smaller than the sample
size. In the limit of small amplitude ac excitation fields ($0.02\,$Oe in our
case), the non-dissipative part Re$\,\chi$ is determined by the ac field
penetration depth $\delta=(\lambda^{2}+\lambda_{C}^{2})^{1/2}$, where
$\lambda$ is the London penetration depth and the Campbell length $\lambda
_{C}$ characterizes the pinning and the VL elastic response \cite{Kosh,Brandt}%
. For $f\sim10\,$MHz, the skin-depth due to the normal excitations vastly
exceeds the size of our samples and has no effect on $\delta$. In this work we
utilize the sensitivity of this method to small changes in $\lambda_{C}$ that
accompany the VL transitions to map the high-field part of the $H_{\Box}(T)$
line unattainable by other techniques.

LuNi$_{2}$B$_{2}$C (Lu1221) were grown out of Ni$_{2}$B flux
\cite{cheon,Gammel}. Powder X-ray diffraction spectra indicate that there were
no detectable second phases present. YNi$_{2}$B$_{2}$C was grown using
vertical zone melting method with a commercial $4$-mirrors image furnace
(model FZ-T-4000-H-VI-VPM-PC, Crystal Systems Corp., Japan). We have studied
both as grown and annealed in active vacuum (950$^{o}$C, 48 h) samples
\cite{miao} of a typical size $1\times1\times0.1$ mm$^{3}$. Residual
resistance ratios (RRR) of Lu-1221 and Y-1221 were about $25$, the critical
temperatures $T_{c}\approx16.2$ and $15.2$ K and the upper critical fields
$H_{c2}\approx9.2$ and $8.8\,$ T, respectively. $H_{c2}$ was determined as the
onset point of the diamagnetic signal. Similar values were obtained from
resistivity onset as well as extrapolation technique used in Ref.\cite{cheon}.

\begin{figure}
[tbh]
\begin{center}
\includegraphics[
height=6.3504cm,
width=8.1012cm
]%
{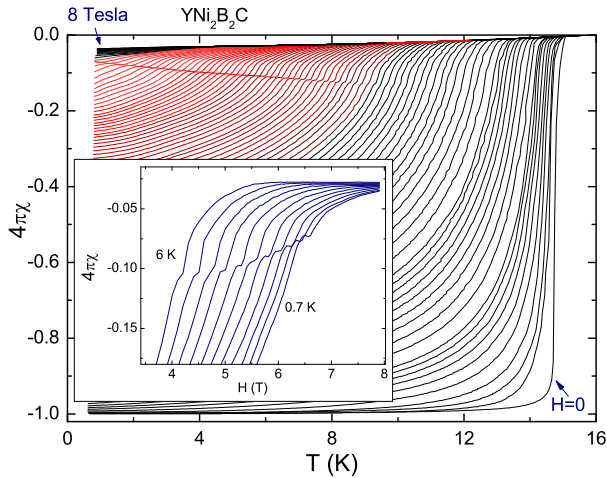}%
\caption{(Color online) $\chi\left(  T\right)  $ in Y1221 measured in fixed
fields with $1$ kOe step. The anomaly corresponding to the VL transition is
clearly seen in the upper left corner (region of red color online).
\textit{Inset:} $\chi\left(  H\right)  $ scans showing the same anomaly.}%
\label{fig1}%
\end{center}
\end{figure}

The temperature scans of the dynamic susceptibility Re$\chi$ at fixed dc
fields from $0$ to $8$\thinspace T for the Y-1221 single crystal with
$H_{dc}||H_{ac}$ along the $c$ axis are shown in Fig.\thinspace\ref{fig1}.
Hereafter we deal only with Re$\chi$ and omit \textquotedblleft Re" for
brevity; - this quantity is normalized so as to have $-1/4\pi$ in the limit
$H_{dc}\rightarrow0$ and to correspond to the full expulsion of the ac signal
for $\delta$ small relative to the sample size. There are clear anomalies in
high-field scans the locus of which appear as a continuous line. The inset to
Fig.\thinspace\ref{fig1} shows magnetic field scans at constant $T$ with a
clearly visible anomaly. The positions $(H,T)$ of the anomaly in field sweeps
coincide with those observed in $\chi\left(  T\right)  $ scans. The field
$H_{c2}(T)$ can also be extracted from the data as points where $\chi$ first
deviates from the upper envelop corresponding to the normal state skin effect;
the observed anomaly clearly differs from $H_{c2}$.%

\begin{figure}
[tbh]
\begin{center}
\includegraphics[
height=9.8914cm,
width=7.0973cm
]%
{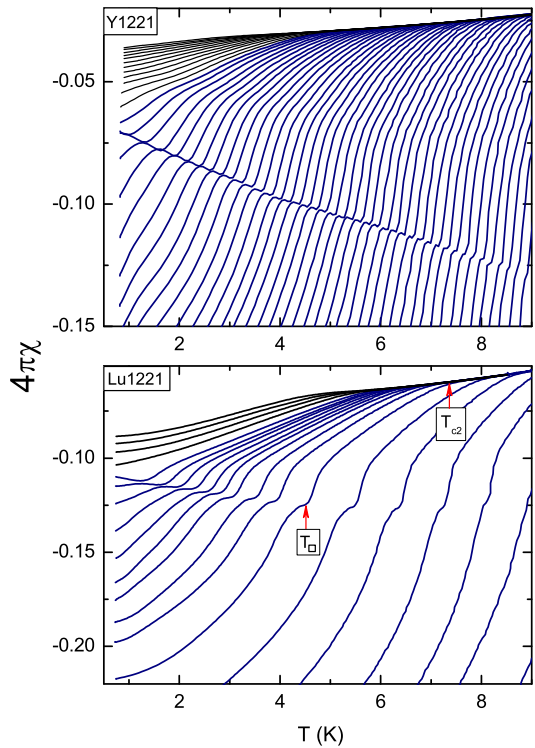}%
\caption{(Color online) Selected areas of $\chi\left(  T,H\right)  $ curves
showing VL transition in Y1221 (top panel) and Lu1221 (bottom panel). At each
fild the characteristic temperatures corrsponding to $H_{c2}\left(  T\right)
$ line as well as $H_{\square}\left(  T\right)  $ are indicated by arrows at
the lower frame.}%
\label{fig2}%
\end{center}
\end{figure}

\begin{figure}
[tbh]
\begin{center}
\includegraphics[
height=6.8974cm,
width=8.1012cm
]%
{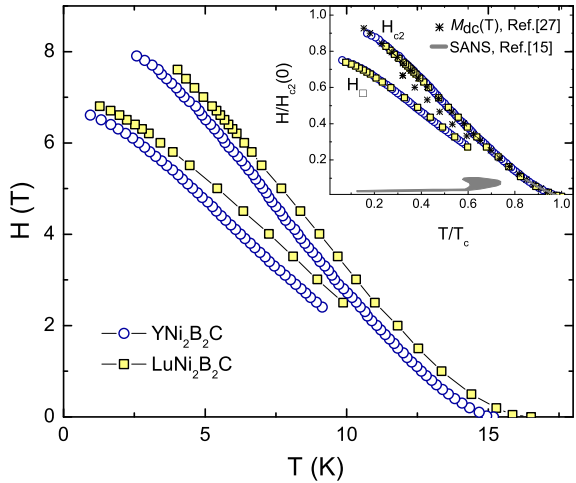}%
\caption{(Color online) Phase diagram obtained from measurements of
$H_{c2}\left(  T\right)  $ and $H_{\square}\left(  T\right)  $. Inset shows
the same data in reduced units. Also shown are the results from Refs.
\cite{park} and \cite{nose}. }%
\label{fig3}%
\end{center}
\end{figure}

Figure \ref{fig2} shows $\chi\left(  T\right)  $ for Y-1221 (upper panel) and
Lu-1221 (lower panel) with a focus on low temperatures and high fields where
the anomaly clearly seen in both samples. Hence, the observed behavior is not
limited to a particular sample, but is seen on both mirror furnace grown
Y-1221 and flux--grown Lu-1221.%

Figure \ref{fig3} is a summary of the data in reduced coordinates $T/T_{c}$
and $H/H_{c2}\left(  0\right)  $. The curves $H_{c2}/H_{c2}(0)$ are nearly the
same for two samples, although as shown in the inset, the absolute values are
somewhat different. For comparison, we also show magnetization data of Ref.
\cite{park} and the low-field branch of the R-S transition according to SANS
measurements of Ref.\cite{nose} (shown with the uncertainty corresponding to
various criteria for the peaks splitting that mark the R-S transition). The
anomaly amplitude decreases with decreasing field and raising temperature
(Fig. \ref{fig2}), and the data cannot be extended to the point where they are
anticipated to merge with the low field and high temperature SANS results.%

The VL response to a small amplitude ac perturbation is determined by the
length \cite{Kosh,Brandt} given by
\begin{equation}
\delta^{2}=\lambda^{2}+\frac{H^{2}}{4\pi K}=\lambda^{2}+\frac{cH\,r_{p}}{4\pi
j_{c}}, \label{Camp}%
\end{equation}
where the term containing the Labusch pinning parameter $K$ is the Campbell's
$\lambda_{C}^{2}$ (in high fields, we do not distinguish between $H$ and the
magnetic induction). In the second formula $\lambda_{C}$ is expressed in terms
of the disorder length $r_{p}$ and the critical current $j_{c}$ for the model
of \textit{weak collective pinning} expected to hold for high quality crystals
of this work \cite{Kosh}. For weak disorder and small $j_{c}$, $\lambda
\ll\lambda_{C}$; also, one can use the relation $r_{p}^{2}=\xi^{2}+\langle
u^{2}\rangle,$ where coherence length $\xi$ is taken to be the range of the
pinning potential and $\langle u^{2}\rangle$ is the mean square thermal
fluctuation of vortex positions \cite{Blatter}.

The VL shear moduli vanish in the limit $H\rightarrow H_{c2}$, the lattice
softens, and $\langle u^{2}\rangle$ diverges. As argued in \cite{GK}, the
nonlocal corrections to the isotropic intervortex interaction responsible for
the square structure are washed away, and the lattice near $H_{c2}$ is always
triangular (for tetragonal crystals in fields along $c$ axis). With decreasing
field, the fluctuations weaken, and at some field $H_{\Box}<H_{c2}$, the
lattice acquires the square structure. The SANS data on cubic V$_{3}$Si
support this picture \cite{Mona}. It has been estimated in
Ref.\,\onlinecite{GK} that this happens when $\langle u^{2}\rangle$ reduces to
values on the order $\xi^{2}$ for clean materials. In most of the domain
$H_{\Box}<H<H_{c2}$, the fluctuations amplitude exceeds $\xi$, and for a
qualitative discussion we can take $r_{p}^{2}\sim\langle u^{2}\rangle$:
\begin{equation}
\delta^{2}\approx\frac{cH}{4\pi j_{c}}\sqrt{\langle u^{2}\rangle}\,.
\label{Camp1}%
\end{equation}

The thermal average $\langle u^{2}\rangle$ depends on \textit{all} VL elastic
moduli. The number of independent moduli and their values depend on the
lattice structure. For fields small on the $H_{c2}$ scale, the moduli were
counted and evaluated in \cite{MK} (and in \cite{Dorsey} within the
Ginzburg-Landau scheme near $T_{c}$). The point relevant for our discussion is
that the number of independent moduli depends on the symmetry of the lattice
(as within the standard elasticity theory for anisotropic solids \cite{LL}).
When the VL rhombic unit cell with diagonals along [100] and [010] of
tetragonal borocarbides (see \cite{boro,ORNL,Grenoble,Argonne,Sakata,nose})
transforms to a square, it acquires an extra symmetry associated with the
four-fold rotational axis. This reduces the number of independent shear moduli
because the shears polarized along [100] and [010] (which are different in the
rhombic lattice) become identical. In addition, the squash modulus for the
deformation responsible for the transition turns zero at the transition point
\cite{MK,Dorsey}. One, therefore, expects $\langle u^{2}\rangle$ be different
on two sides of the transition point.

The calculation of this difference at the transition in high fields would have
required evaluation of the elastic moduli in the presence of thermal
fluctuations, a non-trivial problem to be addressed elsewhere. Still, we can
state that, with increasing scattering, the domain of the square lattice
should shrink and the R-S phase boundary should move to lower reduced fields
$H_{\Box}/H_{c2}$ \cite{GK}. Physically, this is because the transition to the
square structure is caused by the tetragonal symmetry of the Fermi surface and
of the order parameter which enter the intervortex interaction through
nonlocal corrections to the London theory \cite{cubic}. In low fields, the
corrections add the square symmetric interaction at distances on the order of
the nonlocality range $\rho$ which for clean superconductors is on the order
of $\xi_{0}$, zero temperature coherence length. This range, however, shrinks
when the mean-free path becomes shorter. Hence, one has to go to higher fields
for the square-symmetric interaction to cause the R-S transformation. In the
high field domain with fluctuations induced lattice disorder, one has to go to
\textit{lower} fields with weaker fluctuations to reach the domain with
sufficiently strong nonlocal effects to cause the R-S transition. This happens
when $\rho\sim\langle u^{2}\rangle$. Since $\rho$ is reduced by scattering,
the transition takes place at smaller $\langle u^{2}\rangle$, i.e., at lower fields.%

\begin{figure}
[tbh]
\begin{center}
\includegraphics[
height=6.535cm,
width=7.5981cm
]%
{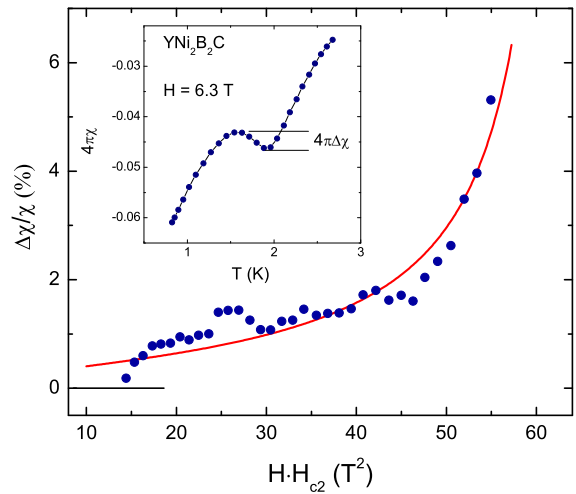}%
\caption{(Color online) Relative jump in $\chi$ at the anomaly in YNi$_{2}%
$B$_{2}$C as a function of $HH_{c2}\left(  T\right)  $. Solid line is a fit to
Eq.(\ref{jump}). \textit{Inset}: the definition of the anomaly size,
$4\pi\Delta\chi$.}%
\label{fig4}%
\end{center}
\end{figure}

To estimate the dynamic susceptibility we note that in our data $-4\pi\chi
\sim0.1$ at the location of the transition. This implies that in our case
$\delta$ is comparable or larger than the sample size. A weak screening of the
ac field in a disc-shaped sample of a radius $R$ is described by the
London-like equation (see, e.g., \cite{Brandt}) $\mathbf{A}-\delta^{2}%
\nabla^{2}\mathbf{A}=0$ for the vector potential $\mathbf{A}=\mathbf{h}%
\times\mathbf{r}/2+\mathbf{a}$, where $\mathbf{a}$ is a small correction to
the uniform field. This gives
\begin{equation}
\mathbf{h}\times\mathbf{r}/2\approx\lambda^{2}\nabla^{2}\mathbf{a=-}%
4\pi\mathbf{\delta^{2}j/c\,.}%
\end{equation}
Given the current $\mathbf{j}$, one readily obtains the magnetic moment and
$\chi=-R^{2}/32\pi\delta^{2}$. Hence,
\begin{equation}
-4\pi\chi=\frac{R^{2}}{8\delta^{2}}=\frac{\pi R^{2}j_{c}}{2cH}\langle
u^{2}\rangle^{-1/2}.
\end{equation}

If $\langle u^{2}\rangle$ varies by $\Delta\langle u^{2}\rangle$, the change
in the measured quantity is:
\begin{equation}
\Delta(-4\pi\chi)=-\frac{\pi R^{2}j_{c}}{4cH\langle u^{2}\rangle^{1/2}}%
\frac{\Delta\langle u^{2}\rangle}{\langle u^{2}\rangle}.
\end{equation}
Since $j_{c}/\langle u^{2}\rangle^{1/2}$ decreases on warming, one expects the
anomaly size to decrease with raising $T$ seen in our data. The relative
change in the susceptibility is:
\begin{equation}
\frac{\Delta\chi}{\chi}=-\frac{\Delta\langle u^{2}\rangle}{2\langle
u^{2}\rangle}\,.
\end{equation}
At the transition $\langle u^{2}\rangle\sim\xi_{0}$ for clean materials. As is
seen at Fig.\ref{fig4}, $\Delta\chi/\chi\sim10^{-2}$ in large fields. Hence,
our method is sensitive to small changes in thermal averages: $\sqrt
{\Delta\langle u^{2}\rangle}\sim\xi_{0}/10$.

The elastic properties of the incompressible square lattice in fixed fields
are characterized by four independent moduli which can be chosen as the shear
$c_{x}$ along one of the square diagonals $x\equiv\,\,$[100], the squash
$c_{s}$ transforming the square to a rhombus, the rotation $c_{r}$, and the
tilt $c_{44}$. At the transition $c_{s}=0$, whereas according to \cite{GK},
\begin{equation}
\frac{c_{r}}{c_{x}}\approx\frac{32\pi\rho^{2}\sqrt{HH_{c2}}}{\phi_{0}}%
\sim\frac{\sqrt{HH_{c2}(T)}}{H_{c2}(0)}\,,
\end{equation}
($\rho$ varies from $0.67\xi_{0}$ to $0.3\xi_{0}$ on warming from 0 to $T_{c}$
in the clean limit). In this situation, Ref.\thinspace\cite{GK} gives $\langle
u^{2}\rangle\approx f(H,T)\ln(c_{x}/c_{r})$, where $f$ does not depend on
elastic moduli. As is shown in \cite{GK,Dorsey}, only the shear moduli
$c_{66}$ change sharply at the transition, so that we estimate:
\begin{equation}
\frac{\Delta\langle u^{2}\rangle}{\langle u^{2}\rangle}\sim\frac{\Delta
c_{66}}{c_{66}\ln(c_{66}/c_{r})}\,
\end{equation}
The ratio $\Delta c_{66}/c_{66}$ should not depend strongly either on $H$ or
on $T$ since the change of the shear moduli is of a geometric nature (due to
the symmetry change). We thus expect the following behavior of the measured
quantity:
\begin{equation}
-\frac{\Delta\chi}{\chi}\approx\frac{C}{\ln[H_{c2}^{2}(0)/HH_{c2}(T)]}
\label{jump}%
\end{equation}
with $C\approx\mathrm{const}$. Fig.\thinspace\ref{fig4} shows the data
$\Delta\chi/\chi$ plotted against $HH_{c2}(T)$; the solid line is calculated
taking $H_{c2}(0)=8.1\,$T with the fit parameter $C=0.75$. A qualitative
nature of our argument notwithstanding, we consider this result as supporting
our interpretation of the $\chi$ anomaly as caused by the R-S structural
transformation of the vortex lattice. It should be noted, that the qualitative
model we use disregards pinning relative to the thermal fluctuations: the
disorder length $r_{p}=\sqrt{\xi^{2}+\langle u^{2}\rangle}\approx\langle
u^{2}\rangle$. In the sample with $9\%$ Co, this is unlikely to be true. This
might be the reason why the anomaly $\Delta\chi/\chi$ is nearly washed away in
this sample. Large scale inhomogeneities may have the same effect.

A similar claim has been made in Ref.\thinspace\cite{park} in which an unusual
behavior of the reversible magnetization $M_{dc}(T)$ was interpreted as caused
by the R-S transition. The position of the anomaly on the $HT$ phase diagram
reported here differs from that of Ref.\thinspace\cite{park}. We did not find
an anomaly in $M_{dc}(T,H)$ in our samples.

To conclude, the tunnel-diode resonator is proven to have enough sensitivity
to measure small changes in the rf penetration depth in the high-field mixed
state (i.e. of the Campbell length) which accompany the VL structural
transformations. The high-field branch of the rhombic-square lattice
transition caused by thermal fluctuation of vortices is mapped up.

We thank J. Yan and G. Lapertot who grew Y1221 crystals in optical furnace.
Work at the Ames Laboratory was supported by the Department of Energy-Basic
Energy Sciences under Contract No. DE-AC02-07CH11358. R. P. acknowledges
support from the NSF grant number DMR-05-53285 and the Alfred P. Sloan Foundation.

\end{document}